\documentclass[page-classic]{epl2}

\title{Weak-interaction mediated rates on iron isotopes for presupernova evolution
of massive stars}
\shorttitle{Weak-interaction mediated rates on iron isotopes for
presupernova evolution}
\author{Jameel-Un Nabi\inst{1,2}}
\institute{
  \inst{1} Faculty of Engineering
Sciences, GIK Institute of Engineering Sciences and Technology,
Topi 23640, Swabi, NWFP, Pakistan\\
  \inst{2} The Abdus Salam ICTP,
Strada Costiera 11, 34014, Trieste, Italy }

\abstract{During the presupernova evolution of massive stars, the
isotopes of iron, $^{54,55,56}$Fe, are advocated to play a key role
inside the cores primarily decreasing the electron-to-baryon ratio
($Y_{e}$) mainly via electron capture processes thereby reducing the
pressure support. Electron decay and positron capture on $^{55}$Fe,
on the other hand, also has a consequential role in increasing the
lepton ratio during the silicon burning phases of massive stars. The
neutrinos and antineutrinos produced, as a result of these
weak-interaction reactions, are transparent to the stellar matter
and assist in cooling the core thereby reducing the entropy. The
structure of the presupernova star is altered both by the changes in
$Y_{e}$ and the entropy of the core material. The aim of this paper
is to report the improved microscopic calculation of Gamow-Teller
(GT$_{\pm}$) strength distributions of these key isotopes of iron
using the pn-QRPA theory. The main improvement comes from the
incorporation of experimental deformation values for these nuclei.
Additionally six different weak-interaction rates, namely electron
\& positron capture, electron \& positron decay, and, neutrino \&
antineutrino cooling rates, were also calculated in presupernova
matter. The calculated electron capture and neutrino cooling rates
due to isotopes of iron are in good agreement with the large-scale
shell model (LSSM) results. The calculated beta decay rates,
however, are suppressed by three to five orders of magnitude.}
\begin{document}

\maketitle

All elements heavier than boron are formed by nuclear reactions
inside stars. Core-collapse supernovae are considered to be one of
the major contributors to the production of elements in the
universe. However, core-collapse simulators, to date, find it
challenging to successfully transform the collapse into an
explosion. The details of the micro-physics in the prevailing
extreme environment is poorly understood.  This include but are
not limited to the explosion mechanism, role of neutrino
interactions, equation of state and the treatment of hydrodynamic
instabilities in three dimensional simulations. A mechanism
involving transfer of energy from neutrinos has long been favored
but other vistas involving rapid rotation and magnetic fields are
also being explored. Whereas in the prevailing environment
reactions mediated by strong and electromagnetic force are in
chemical equilibrium those mediated by the weak interactions are
not (the non-thermal neutrinos produced are transparent to the
stellar matter for densities up to around 10$^{11} gcm^{-3}$) and
have a decisive role to play in the intricate dynamics of
core-collapse. The weak interaction have several crucial effects
in the course of development of a star. These include initiation
of the gravitational collapse of the core of massive stars,
neutronisation of the core material and formation of heavy
elements above iron via the r-process.

In order to understand the supernova explosion mechanism and other
related astrophysical events of significance international
collaborations of astronomers and physicists are being sought. A
large amount of useful observational data is now accessible due to
space- and ground-based telescopes. Considerable efforts are also
being done to determine the equation of state of ultradense
matter. State-of-the-art supernova models employ multidimensional
modelling making use of sophisticated numerical treatments. This
in turn is also made possible by faster computing machines.
Reliable determinations of nuclear properties of many nuclei
(including neutron-rich unstable nuclide) are becoming available
thanks to improved nuclear-mass models coupled with new
experimental measurements.

It is highly desirable to calculate the presupernova stellar
structure with the most reliable physical data and inputs. The
structure of the presupernova star is altered both by the change
in $Y_{e}$ and entropy in its interior. A smaller precollapse iron
core mass and a lower entropy should favor an explosion. A smaller
iron core size implies less energy loss by the shock in
photodisintegrating the iron nuclei in the overlying onion-like
structure whereas a lower entropy environment can assist to
achieve higher densities for the ensuing collapse generating a
stronger bounce and in turn forming a more energetic shock wave
\cite{Bet79}. A smaller entropy can also assist in achieving a
higher final value of $Y_{e}$ by reducing the abundance of free
protons (which act as a major source of electron sink through
electron capture reactions). The entropy profile determines the
extent of the convective burning shells and has significant
effects on presupernova core structure and nucleosynthesis. As
such it is desirable to turn on the nuclear weak reaction network
as early as possible in the simulation codes and dynamically
couple the network with the evolution. A sufficiently detailed and
reliable nuclear reaction network (microscopically calculated
taking into account the nuclear structure details of the
individual nuclei) is more likely to achieve positive results.

Electron/positron captures and $\beta^{\pm}$-decay rates are
amongst the most important nuclear physics inputs that determine
both the $Y_{e}$ and the entropy at the presupernova stage.
Electron capture decreases the number of electrons available for
pressure support whereas beta decay acts in the opposite
direction. Both processes directly affect the overall
lepton-to-baryon ratio of the core. The neutrinos and
antineutrinos produced as a result of these nuclear weak reactions
are transparent to the stellar matter at presupernova densities
and therefore assist in cooling the core to a lower entropy state.
These weak-interaction rates are required not only in the accurate
determination of the structure of the stellar core but also bear
significance in (explosive) nucleosynthesis and element abundance
calculations. Weak interactions in presupernova stars are known to
be dominated by allowed Fermi and Gamow-Teller transitions
\cite{Bet79}. In particular, Gamow-Teller (GT) properties of
nuclei in the region of medium masses around A=56 are of special
importance because they are the main constituents of the stellar
core in presupernova conditions.

The first ever extensive calculation of stellar weak-interaction
rates was performed by Fuller, Fowler and Newmann \cite{Ful80}.
These pioneering calculations included rates for electron/positron
capture, electron/positron decay and associated (anti)neutrino
energy losses for around 226 nuclei with masses between A = 21 and
60. The authors also incorporated all experimental data available
at that time. The GT strength distributions and excitation
energies were calculated using a zero-order shell model.
Aufderheide and collaborators \cite{Auf94} stressed on the
importance of beta decay in the iron core just prior to the
collapse of the core and extended the calculations \cite{Ful80}
for heavier nuclei with A $>$ 60. Later experimental results
\cite{Roe93,And90,Elk94,Rap83} revealed the misplacement of the GT
centroid adopted in the parameterizations of Ref. \cite{Ful80} and
subsequently used in the calculation by Ref. \cite{Auf94}. Since
then considerable efforts were made on the microscopic calculation
of these weak-interaction rates. Large-scale shell model
(hereafter LSSM)(e.g. \cite{Lan00}) and the proton-neutron
quasiparticle random phase approximation (hereafter pn-QRPA)
theory (e.g. \cite{Nab99}) were two of the most successful and
extensively used models for the microscopic calculation of stellar
weak rates.

The pn-QRPA theory is an efficient way to generate GT strength
distributions. These strength distributions constitute a primary
and nontrivial contribution to the capture and decay rates among
iron-regime nuclide. Nabi and Klapdor-Kleingrothaus used the
pn-QRPA theory, for the first time, to calculate the
weak-interaction rates over a wide range of temperature and
density scale for sd- \cite{Nab99} and fp/fpg-shell nuclei
\cite{Nab04} in stellar matter. Since then these calculations were
refined and further improved with use of more efficient
algorithms, computing power, incorporation of latest data from
mass compilations and experimental values, and fine-tuning of
model parameters (e.g. Refs. \cite{Nab07a,Nab07b,Nab08a}). The
reliability of pn-QRPA calculation was established and discussed
in detail in Ref. \cite{Nab04}. The associated uncertainties
involved in the calculations were highlighted in Ref.
\cite{Nab08b}. In this paper I use this established and improved
model (as also explained later) to calculate the weak-interaction
rates for key iron isotopes in presupernova conditions.

The weak decay rate from the $\mathit{i}$th state of the parent to
the $\mathit{j}$th state of the daughter nucleus is given by
\begin{equation}
\label{eq.1} \lambda_{ij} =ln2
\frac{f_{ij}(T,\rho,E_{f})}{(ft)_{ij}},
\end{equation}
where $(f_{ij})$ are the phase space integrals and are functions
of temperature ($T$), density ($\rho$) and Fermi energy ($E_{f}$).
The pn-QRPA model is employed to calculate the $(ft)_{ij}$ values
for the transitions which are directly related to the reduced
transition probabilities of the Fermi and GT transitions. The
total weak-interaction rate per unit time per nucleus is finally
given by performing a double summation of the type
\begin{equation}
\label{eq.2} \lambda =\sum _{ij}P_{i} \lambda _{ij}.
\end{equation}
The summation was carried out over all parent and daughter states
until satisfactory convergence in the rate calculation was
achieved. Here $P_{i}$ is the probability of occupation of parent
excited states and follows the normal Boltzmann distribution. The
detailed formalism may be found in Ref. \cite{Nab04}. The pn-QRPA
theory allows a microscopic state-by-state calculation of both
sums present in eq.~(\ref{eq.2}). In other words the pn-QRPA model
calculates the GT strength distribution of all parent excited
states in a microscopic fashion. This salient feature of the
pn-QRPA model greatly increases the reliability of the calculated
rates in stellar matter where there exists a finite probability of
occupation of parent excited states. Other models revert to
approximations like 'Brink's hypothesis' (in electron capture
direction) and 'back resonances' (in $\beta$-decay direction) in
order to perform these summations. Brink's hypothesis states that
GT strength distribution on excited states is \textit{identical}
to that from ground state, shifted \textit{only} by the excitation
energy of the state. GT back resonances are the states reached by
the strong GT transitions in the inverse process (electron
capture) built on ground and excited states.

Three key isotopes of iron, $^{54,55,56}$Fe, were chosen for the
calculation of GT$_{\pm}$ strength distributions and associated weak
rates using the pn-QRPA model. Reasonable experimental data were
available for the even-even isotopes of iron ($^{54,56}$Fe) to test
the model. Low-lying parent excited states of iron isotopes
(specially  $^{55}$Fe) have a finite probability of occupation under
stellar conditions and a microscopic calculation of GT$_{\pm}$
strengths from these excited states is desirable. The pn-QRPA model
is specially designed to perform this task. Aufderheide and
collaborators \cite{Auf94} ranked $^{54,55,56}$Fe amongst the most
influential nuclei with respect to their importance for the electron
capture process for the early presupernova collapse. Later Heger et
al. \cite{Heg01} studied the presupernova evolution of massive stars
and rated $^{54,55,56}$Fe amongst the very important nuclei
considered to be most important for decreasing $Y_{e}$ during the
oxygen and silicon burning phases. Besides, $^{55}$Fe was also found
to be in the top five list of nuclei that increases $Y_{e}$ via
positron capture and electron decay during the silicon burning
phases \cite{Heg01}. The rationale behind this project is to present
an alternate microscopic and accurate estimate of weak-interaction
mediated rates on iron isotopes, on a detailed temperature-density
grid suitable for interpolation purposes, for the collapse
simulators which may be used as a reliable source of nuclear physics
input in the simulation codes.

The pn-QRPA calculation performed in this project was further
improved by an optimum choice of model parameters and
incorporation of latest experimental data. Recent study by Stetcu
and Johnson \cite{Ste04} stressed particularly on the importance
of deformation parameter in the QRPA model for improved results.
Rather than incorporating deformations calculated from some
theoretical mass model (as used in earlier calculations of pn-QRPA
rates \cite{Nab04,Nab99}), for the first time the experimentally
adopted value of the deformation parameter for $^{54,56}$Fe, taken
from Raman et al. \cite{Ram87}, was employed in the calculation.
The better choice of model parameters resulted in an improved
agreement of the calculated centroids of the GT strength
distributions with measurements. The GT$_{+}$ centroid for
$^{54}$Fe ($^{56}$Fe) was improved from 6.29 MeV (4.68 MeV) to
4.06 MeV (3.13 MeV). The corresponding measured centroids are 3.7
$\pm$ 0.2 MeV (2.9 $\pm$ 0.2 MeV). The total calculated strengths
$\Sigma S_{\beta^{\pm}}$ were also in better agreement with the
measured values (see Table 1). For the case of $^{55}$Fe (where
measurement lacks) the deformation of the nucleus was calculated
using the mass compilation of M\"{o}ller and Nix \cite{Moe81}. In
order to further increase the reliability of the calculated weak
rates experimental data were incorporated in the calculation
wherever possible. For details see Ref.~\cite{Nab08a}. A
state-by-state calculation of GT$_{\pm}$ strength was performed
for a total of 246 parent excited states in $^{54}$Fe, 297 states
in $^{55}$Fe and 266 states in $^{56}$Fe. For each parent excited
state, transitions were calculated for a total of 150 daughter
excited states. The band width of energy states was chosen
accordingly to cover an excitation energy range of (15 - 20) MeV
in parent and daughter. The summations in eq.~(\ref{eq.2}) were
done to ensure satisfactory convergence. The use of a separable
interaction assisted in the incorporation of a luxurious model
space of up to 7 major oscillator shells which in turn made
possible to consider these many excited states both in parent and
daughter nuclei (interested readers are referred to
Ref.~\cite{Nab04} for details of the model description).

Both $\beta$-decay and capture rates are very sensitive to the
location of the GT$_{+}$ centroid \cite{Auf96}. An overall quenching
factor of 0.6 \cite{Gaa83} was adopted in the current pn-QRPA
calculation of GT strength in both directions for all iron isotopes.
A considerable amount of uncertainty is present in the calculations
and measurements of GT strength distribution. The associated
uncertainties in the pn-QRPA model was discussed in Ref.
\cite{Nab08b}.  Table~\ref{tab.1} displays the comparison of the
calculated GT centroids and total strengths ($\Sigma
S_{\beta^{\pm}}$) against measurement, LSSM calculation and earlier
pn-QRPA calculation of Ref. \cite{Nab04} for $^{54,56}$Fe. For the
case of $^{54}$Fe ($^{56}$Fe) the value of total $S_{\beta^{+}}$ was
taken from Ref. \cite{Roe93} (Ref. \cite{Elk94}) whereas the total
measured $S_{\beta^{-}}$ strength was taken from Ref. \cite{And90}
(Ref. \cite{Rap83}). The pn-QRPA calculated total strengths, $\Sigma
S_{\beta^{+}}$, lie closer to the upper bound of the measured
values. The calculated $\Sigma S_{\beta^{-}}$ strengths match very
well with the measurements. The calculated weak rates are sensitive
to the location of GT$_{+}$ centroids \cite{Auf96}. The pn-QRPA
centroids are within the upper range of the measured values and are
placed at relatively higher energies in daughter nuclei as compared
to LSSM numbers. Experimental (p,n) data are also available for
$^{54,56}$Fe. According to Anderson and collaborators \cite{And90}
the amount of GT strength in the background and continuum remained
highly uncertain. For $^{54}$Fe ($^{56}$Fe) the centroid of the data
of discrete peaks \cite{And90} (\cite{Rap83}) was calculated to be
7.63 MeV (8.27 MeV). The pn-QRPA calculated values are 5.08 MeV and
5.61 MeV, respectively. Despite greater experimental certainties in
the (p,n) data one notes that the pn-QRPA places the centroid at
much lower energies in daughter nuclei as compared to measurements.
However it is to be noted that the total strengths (in both
direction) compare very well with the measured values. Also the
location of GT$_{+}$ centroids is in very good agreement with the
measured data which is mainly responsible for the calculation of
weak rates on iron isotopes \cite{Auf96}. The Ikeda sum rule was
satisfied for the ground states of iron isotopes for the unquenched
values in the calculation for the complete range of excitation
energies covered in daughter nuclei as discussed above. (Note that
the values of the total $S_{\beta^{\pm}}$ strengths given later in
Table 1 and Table 2 are calculated up to an excitation energy of
12MeV in daughter.) For reasons not known to the author, LSSM did
not present the location of GT$_{-}$ centroids in their paper
\cite{Lan00}. Table 1 also highlights the improvement in the current
calculation as compared to the pn-QRPA calculation of Ref.
\cite{Nab04}.

Table~\ref{tab.2} basically shows that the Brink's hypothesis
(also used in the LSSM calculation) may not be a good
approximation to use in calculation of stellar weak rates for iron
isotopes. The energies in parenthesis are experimental. The
centroids of the GT$_{\pm}$ strength distributions are shifted to
higher excitation energies in daughter nuclei for the parent
excited states as compared to the ground state centroids. The
total strengths $S_{\beta^{\pm}}$ also change appreciably for the
excited states. These changes have an effect on the calculated
weak rates specially at low temperatures and densities (where
specific low-lying transitions may dominate the rate calculation).

\begin{figure}
\onefigure[width=1.2\textwidth]{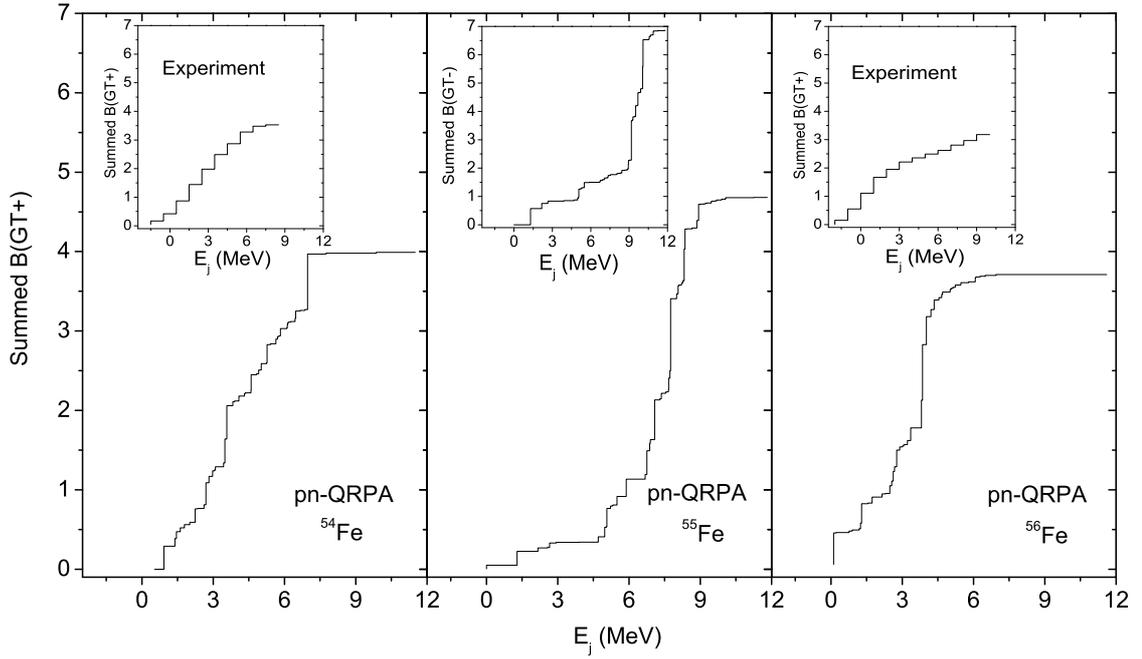} \caption{Cumulative
Gamow-Teller (GT$_{+}$) strength distributions for iron isotopes.
The inset depicts experimental results. For $^{55}$Fe (middle
panel) the inset shows the summed B(GT$_{-}$) strength
distribution.} \label{fig.1}
\end{figure}

The cumulative GT$_{+}$ strength distributions for the iron
isotopes are presented in fig.~\ref{fig.1}. The insets show the
corresponding experimental results. For the case of $^{55}$Fe
where no measurement is available, the summed GT$_{-}$ strength
distribution is shown instead in the inset. As can be seen from
fig.~\ref{fig.1}, the calculated distributions are fragmented,
well spread and are in good agreement with the measured
distributions.

\begin{figure}
\onefigure{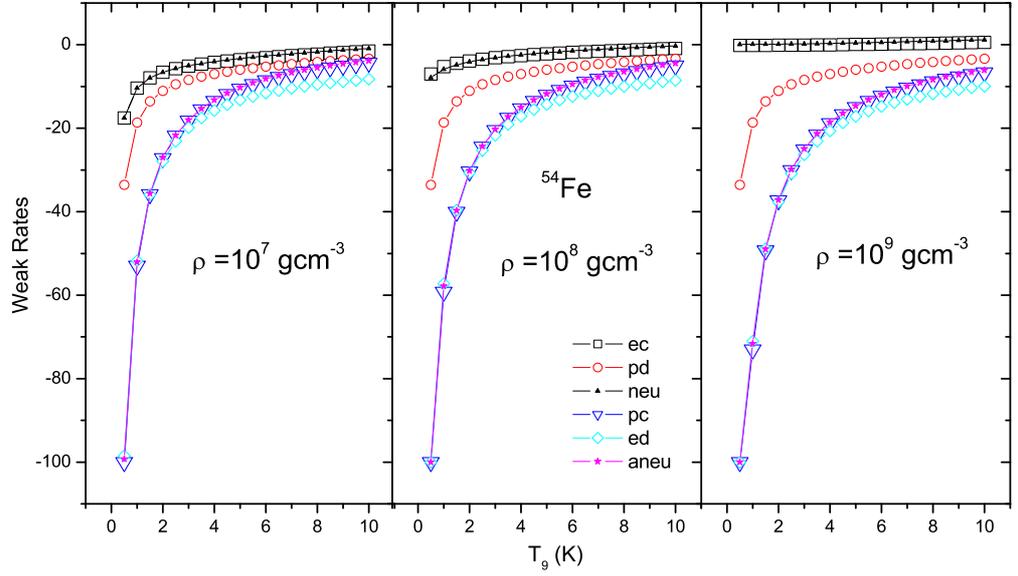} \caption{(Color online) Weak-interaction rates
for $^{54}$Fe during presupernova evolution of massive stars
calculated using the pn-QRPA theory. Here ec, pd, neu, pc, ed and
aneu stand for calculated electron capture, positron decay,
neutrino energy loss, positron capture, electron decay and
antineutrino energy loss rates, respectively. For units of
calculated weak rates see text.} \label{fig.2}
\end{figure}

\begin{figure}
\onefigure{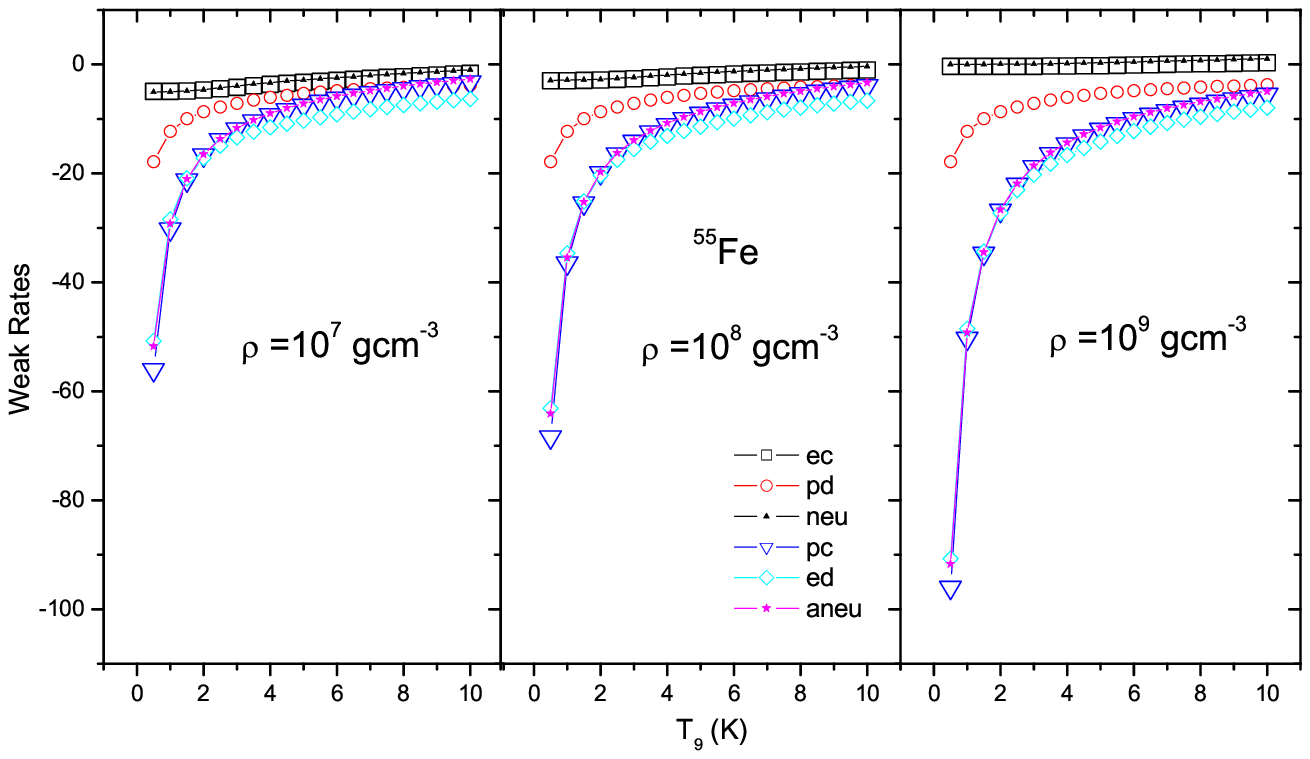} \caption{(Color online) Same as fig.~\ref{fig.2}
but for $^{55}$Fe.} \label{fig.3}
\end{figure}

\begin{figure}
\onefigure{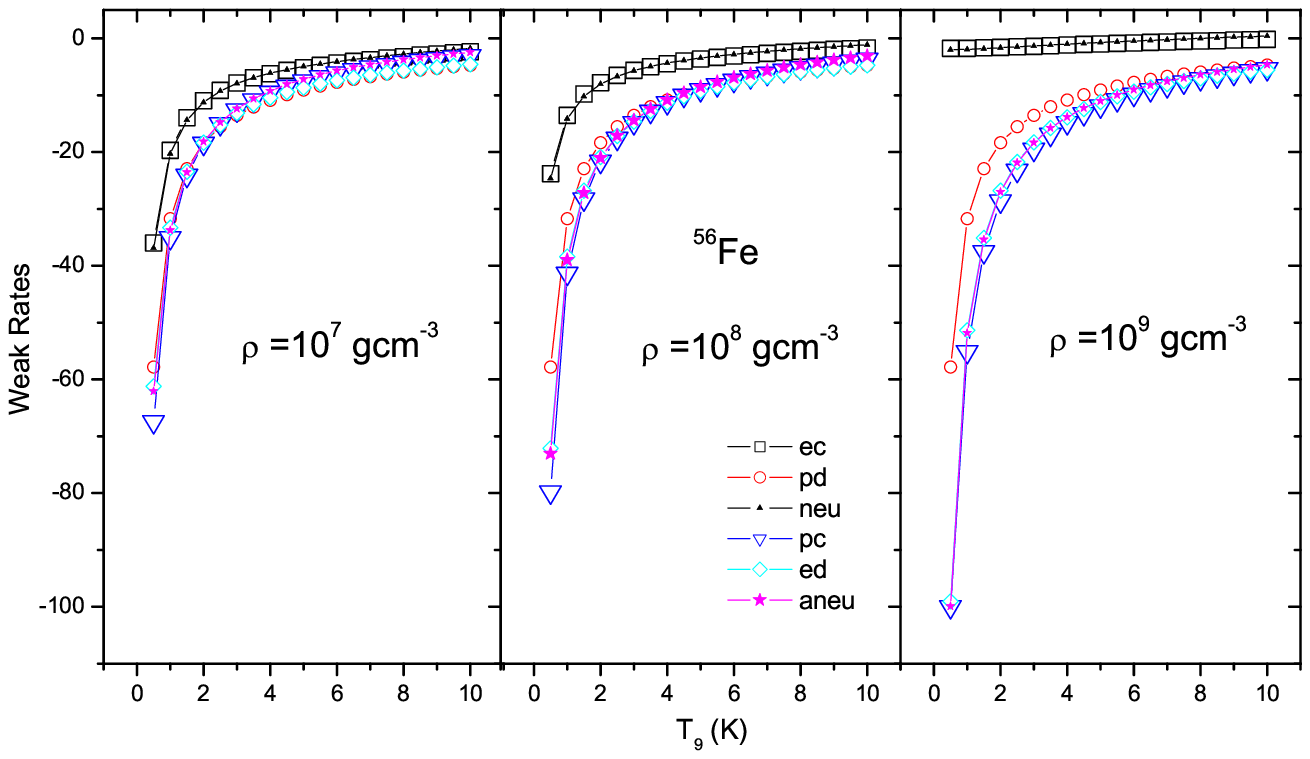} \caption{(Color online) Same as fig.~\ref{fig.2}
but for $^{56}$Fe.} \label{fig.4}
\end{figure}

The calculated weak rates for the iron isotopes during
presupernova evolution from post-oxygen burning till the
presupernova model for massive stars are depicted in
figs.~\ref{fig.2}, \ref{fig.3} and \ref{fig.4} (color online).
Each figure consists of three panels of increasing presupernova
density and depicts six weak-interaction rates calculated using
the pn-QRPA model at astrophysically relevant temperature and
density scales. These include electron capture (ec), positron
decay (pd), neutrino energy losses (neu), positron capture (pc),
electron decay (ed) and antineutrino energy loss rates (aneu) for
$^{54,55,56}$Fe. All weak rates are given in log to base 10
scales. The capture and decay rates (open markers) are given in
units of $s^{-1}$ whereas the neutrino and antineutrino energy
losses (filled upper triangle and star, respectively) are given in
units of $MeV s^{-1}$. In these figures $T_{9}$ gives the stellar
temperature in units of $10^{9}$ K. It may be noted from these
figures that electron capture, positron decay and neutrino energy
losses are more dominant than the other three weak-interaction
processes specially at low temperatures. Furthermore the electron
capture rates and neutrino energy losses on one hand and the
electron decay (also positron capture) rates and antineutrino
energy losses on the other hand are fairly similar in magnitude
during the presupernova evolution of massive stars. These figures
also imply that during this time period the electron capture and
neutrino cooling rates are the dominant weak-processes and hence
most important for core-collapse simulators.

How does the pn-QRPA calculated rates compare with the LSSM results
during the presupernova evolution of massive stars? The answer is
given in tables~\ref{tab.3}, \ref{tab.4}. These tables show the
comparison of weak rate calculations in astrophysically relevant
regions (as determined by the studies of presupernova evolution of
massive stars by Ref. \cite{Heg01}). In these tables $T_{9}$ gives
the stellar temperature in units of $10^{9}$ K and $\rho_{7}$ is the
density in units of $10^{7} g cm^{-3}$. In table~\ref{tab.3},
$R_{ec}$ and $R_{pd}$ are the ratios of pn-QRPA calculated electron
capture and positron decay rates to those calculated using LSSM (in
this direction Fe isotopes transform to Mn). As discussed above the
electron capture rates are much bigger (by many orders of magnitude)
than the competing positron decay rates and hence are more important
for core-collapse simulators. The neutrino produced as a result of
these two weak processes are transparent to the stellar matter
during the presupernova evolution of massive stars and their cooling
capability is determined by the neutrino energy loss rates and
$R_{\nu}$ gives the corresponding pn-QRPA to LSSM calculated ratio.
For all isotopes of iron the pn-QRPA calculated electron capture and
neutrino energy losses are in very good agreement with the LSSM
rates. Positron decay rates are much smaller and relatively less
important for simulation codes. The comparison of positron decay
rates of $^{54,55}$Fe is very good at lower temperatures ($T_{9}$ =
2--3). Otherwise the pn-QRPA calculated positron decay rates are
smaller by around one to two orders of magnitude.

Table~\ref{tab.4} shows the ratios of the reported positron capture
($R_{pc}$), electron decay ($R_{ed}$) and antineutrino energy loss
rates ($R_{\bar{\nu}}$) to LSSM rates for $^{55}$Fe during the
silicon burning phases of massive stars. Positron capture and
electron decay rates are of comparable magnitudes and compete well
with each other (see figures~\ref{fig.2},\ref{fig.3},\ref{fig.4}).
The positron capture rates on $^{55}$Fe are in reasonable agreement
with the LSSM results whereas the calculated $\beta$-decay rates are
smaller by three to five orders of magnitude. It was pointed out by
authors in \cite{Auf94,Heg01} that $\beta$-decay rates are
negligible in the presupernova model. During the silicon burning
phases the pn-QRPA calculated antineutrino energy loss rates are
also smaller by up to five orders of magnitude. The beta decay rates
are very sensitive to the available phase space as compared to the
capture rates. The phase space is given by ($Q + E_{i} -E_{j}$)
where $E_{i}(E_{j})$ are the parent (daughter) excitation energies
and $Q$ is the Q-value of the reaction. The calculated GT strength
distribution and placement of GT$_{-}$ centroid caused the big
suppression in the pn-QRPA $\beta$-decay rates. The centroid
placement in reported calculation is given in table~\ref{tab.2}. The
$\beta$-decay rates tend to go down at high densities due to the
blocking of the available phase space of increasingly degenerate
electrons. At high densities the contribution to the total
$\beta$-decay rates by the excited states is very important,
specially for higher temperatures, because such states, due to their
higher available phase space, are still unblocked. The enhancement
in LSSM beta decay rates may also be attributed to the inclusion of
back resonances in their calculation. It is to be noted that these
rates are very small numbers and can change by orders of magnitude
by a mere change of 0.5 MeV, or less, in parent or daughter
excitation energies, $E_{i}(E_{j})$, and are more reflective of the
uncertainties in the calculation of energies.

The aim of this paper was to present an alternate microscopic and
accurate estimate of weak-interaction mediated rates on key iron
isotopes for the collapse simulators. As mentioned earlier, the
pn-QRPA model makes a microscopic calculation of GT strength from
\textit{all} parent excited states possible which greatly
increases the utility of this model in calculation of stellar
weak-interaction rates. The complete set of detailed calculation
of all six weak rates for iron isotopes as a function of stellar
temperature, density and Fermi energy, suitable for core-collapse
simulations and interpolation purposes, is available as ASCII
files and can be requested from the author. A detailed study of GT
strength distributions and weak-interaction rates of these iron
isotopes along with a comprehensive comparison with previous
calculations will be presented elsewhere. The overall final lepton
fraction and entropy of the collapsing core can be determined by
the calculation of similar reaction rates for other important
iron-regime nuclei. Work is in progress on improved calculations
of pn-QRPA rates of other astrophysically important fp-shell
nuclide and one can hope that in near future a microscopic,
reliable and complete set of nuclear physics input will be
available for the core-collapse simulators to probe for some
possible interesting outcome.

\acknowledgments The author would like to acknowledge the kind
hospitality provided by the Abdus Salam ICTP, Trieste, where part of
this project was completed. The author further wishes to acknowledge
the comments of one of the Referee which lead to the proper
incorporation of experimental data in case of calculations for
$^{55}$Fe.

\newpage
\begin{table}
\caption{Comparison of measured GT$_{+}$ centroids and total
$S_{\beta^{\pm}}$ strengths with microscopic calculations of the
improved pn-QRPA model (this work), large scale shell model (LSSM)
\cite{Lan00} and those of Ref. \cite{Nab04} in
$^{54}Fe$($^{56}Fe$). For experimental references see text.}
\label{tab.1}
\begin{center}
\scriptsize\begin{tabular}{cccc} Source & E$(GT_{+})$ & $\Sigma
S_{\beta^{-}}$ & $\Sigma S_{\beta^{+}}$
\\\hline
Experiment  & 3.7$\pm0.2$(2.9$\pm0.2$) & 7.5$\pm 0.7$(9.9$\pm2.4)$
&3.5$\pm0.7$(2.9$\pm0.3$)\\
pn-QRPA        & 4.06(3.13) & 7.56(10.74) & 4.26(3.71)\\
LSSM & 3.78(2.60) &7.11(9.80) & 3.56(2.70)\\
Ref. \cite{Nab04} & 6.29(4.68)& 9.47(12.47) & 4.49(4.72)\\
\end{tabular}
\end{center}
\end{table}

\begin{table}
\caption{Comparison of pn-QRPA calculated centroids E(GT$_{\pm}$)
and total $S_{\beta^{\pm}}$ strengths for the ground and first two
excited states of $^{54,55,56}Fe$. The cut-off energy in daughter
nuclei is 12 MeV.} \label{tab.2}
\begin{center}
\scriptsize\begin{tabular}{ccccc}
\\ States &
E$(GT_{+}) [MeV]$ & E$(GT_{-}) [MeV]$ & $\Sigma S_{\beta^{+}}$ &
$\Sigma S_{\beta^{-}}$
\\\hline
$^{54}Fe$ (0.0 MeV)       & 4.06 & 5.08 & 4.26 & 7.56\\
$^{54}Fe$ (1.41 MeV)        & 7.10 & 7.81    & 5.12 &  6.97\\
$^{54}Fe$ (2.56 MeV)        & 7.48 & 8.23    & 3.84  &
9.23\\\hline
$^{55}Fe$ (0.0 MeV)       & 7.12 &  8.28 & 4.68 & 6.87\\
$^{55}Fe$ (0.41 MeV)        & 7.38 & 8.56    & 4.43  & 8.87\\
$^{55}Fe$ (0.93 MeV)        & 7.40 & 8.83    & 5.03  &
6.87\\\hline
$^{56}Fe$ (0.0 MeV)       & 3.13 & 5.61 & 3.71 & 10.74\\
$^{56}Fe$ (0.85 MeV)        & 6.17 & 8.89    & 5.15 & 8.04\\
$^{56}Fe$ (2.08 MeV)        & 6.74&  8.76    & 4.15 & 10.21\\
\end{tabular}
\end{center}
\end{table}

\begin{table}
\caption{Ratios of pn-QRPA weak rates to those calculated using
LSSM \cite{Lan00} in presupernova conditions for key iron
isotopes. See text for explanation of symbols.} \label{tab.3}
\begin{center}
\scriptsize\begin{tabular}{c|ccc|ccc}
 & & $T_{9}=2$ & & & $T_{9} =3$ &
\\\hline
$\mathbf{^{54}Fe}$& $R_{ec}$ & $R_{pd}$ &$R_{\nu}$ & $R_{ec}$ &
$R_{pd}$ &$R_{\nu}$
\\
$\rho_{7} = 1$& 3.26E+00 & 1.33E+00 &    3.59E+00 & 2.38E+00 &
6.37E-01 & 2.47E+00 \\
$\rho_{7} = 10$& 3.40E+00 & 1.33E+00 &    3.23E+00 & 2.26E+00 &
6.37E-01 & 2.27E+00 \\
$\rho_{7} = 100$& 1.17E+00 & 1.33E+00 &    1.41E+00 & 1.11E+00 &
6.37E-01 & 1.31E+00 \\
\hline $\mathbf{^{55}Fe}$ & & & & \\
$\rho_{7} =1$& 1.09E+00 &
1.57E+00 &    1.10E+00 & 1.44E+00 & 1.40E+00 & 1.49E+00 \\
$\rho_{7} = 10$& 1.81E+00 & 1.57E+00 &    1.43E+00 & 1.63E+00 &
1.40E+00 & 1.53E+00 \\
$\rho_{7} = 100$& 1.26E+00 & 1.57E+00 &    1.59E+00 & 1.06E+00 &
1.40E+00 & 1.31E+00 \\
\hline $\mathbf{^{56}Fe}$ & & & & \\
$\rho_{7} =1$& 1.14E+00 &
1.77E-01 &    1.07E+00 & 1.17E+00 & 7.80E-02 & 1.08E+00 \\
$\rho_{7} = 10$& 1.35E+00 & 1.77E-01 &    1.32E+00 & 1.37E+00 &
7.78E-02 & 1.29E+00 \\
$\rho_{7} = 100$& 2.11E+00 & 1.77E-01 &    1.97E+00 & 1.80E+00 &
7.78E-02 & 1.71E+00 \\
&&&&&&\\
& & $T_{9}=5$ & & & $T_{9} =10$ &
\\\hline
$\mathbf{^{54}Fe}$& $R_{ec}$ & $R_{pd}$ &$R_{\nu}$ & $R_{ec}$ &
$R_{pd}$ &$R_{\nu}$
\\$\rho_{7} = 1$& 1.39E+00 & 2.29E-01 &    1.31E+00 & 9.95E-01 &
1.07E-01 & 8.34E-01 \\
$\rho_{7} = 10$& 1.35E+00 & 2.31E-01 &    1.36E+00 & 9.91E-01 &
1.10E-01 & 8.97E-01 \\
$\rho_{7} = 100$& 9.93E-01 & 2.31E-01 &    1.10E+00 & 9.12E-01 &
1.11E-01 & 9.31E-01 \\
\hline $\mathbf{^{55}Fe}$ & & & & \\
$\rho_{7} =1$& 1.75E+00 &
6.43E-01 &    1.79E+00 & 1.12E+00 & 7.36E-02 & 1.02E+00 \\
$\rho_{7} = 10$& 1.58E+00 & 6.44E-01 &    1.72E+00 & 1.09E+00 &
7.76E-02 & 1.07E+00 \\
$\rho_{7} = 100$& 8.75E-01 & 6.44E-01 &    1.12E+00 & 8.55E-01 &
7.91E-02 & 9.89E-01 \\
\hline $\mathbf{^{56}Fe}$ & & & & \\
$\rho_{7} =1$& 1.14E+00 &
4.99E-02 &    1.03E+00 & 1.02E+00 & 1.04E-01 & 8.83E-01 \\
$\rho_{7} = 10$& 1.23E+00 & 5.01E-02 &    1.14E+00 & 1.03E+00 &
1.06E-01 & 9.20E-01 \\
$\rho_{7} = 100$& 1.36E+00 & 5.01E-02 &    1.35E+00 & 1.03E+00 &
1.06E-01 & 1.01E+00 \\
\end{tabular}
\end{center}
\end{table}

\begin{table}
\caption{Ratios of pn-QRPA weak rates to those calculated using
LSSM \cite{Lan00} in presupernova conditions for $^{55}Fe$. See
text for explanation of symbols.} \label{tab.4}
\begin{center}
\scriptsize\begin{tabular}{c|ccc|ccc|ccc}
 & & $T_{9}=2$ & & & $T_{9} =3$ & & & $T_{9} =5$ &
\\\hline
$\mathbf{^{55}Fe}$& $R_{pc}$ & $R_{ed}$ &$R_{\bar{\nu}}$ &
$R_{pc}$ & $R_{ed}$ &$R_{\bar{\nu}}$ & $R_{pc}$ & $R_{ed}$
&$R_{\bar{\nu}}$
\\$\rho_{7} = 1$& 1.52E-01 & 2.47E-03 &    1.45E-02 & 4.72E-01 &
5.92E-04 & 2.89E-02 &    7.01E-01 & 1.54E-04 &
1.22E-01 \\
$\rho_{7} = 10$& 1.52E-01 & 8.51E-05 &    6.37E-04 & 4.72E-01 &
4.61E-05 & 2.61E-03 &    7.00E-01 & 4.61E-05 &
2.08E-02 \\
$\rho_{7} = 100$& 1.52E-01 & 2.11E-05 &    1.79E-04 & 4.71E-01 &
1.27E-05 & 7.96E-04 &    7.00E-01 & 1.79E-05 &
6.58E-03 \\
\end{tabular}
\end{center}
\end{table}

\end{document}